\documentclass[aps,prd,onecolumn,groupedaddress,showpacs,nofootinbib,amssymb]{revtex4}
\usepackage[dvips]{graphicx}
\usepackage{amssymb}
\usepackage{amsmath}
\usepackage{graphicx,,color}
\usepackage{amsfonts}
\usepackage{bm}
\usepackage{cancel}
\usepackage{comment}

\newcommand\be{\begin{equation}}
\newcommand\ee{\end{equation}}

\allowdisplaybreaks[4]

\begin{document}

\tolerance=5000

\title{A Refined Einstein-Gauss-Bonnet Inflationary Theoretical Framework}
\author{V.K.~Oikonomou,$^{1,2}$\,\thanks{v.k.oikonomou1979@gmail.com}\\
$^{1)}$ Department of Physics, Aristotle University of
Thessaloniki, Thessaloniki 54124,
Greece\\
$^{2)}$ Laboratory for Theoretical Cosmology, Tomsk State
University of Control Systems and Radioelectronics, 634050 Tomsk,
Russia (TUSUR)\\}

 \tolerance=5000

\begin{abstract}
We provide a refined and much more simplified
Einstein-Gauss-Bonnet inflationary theoretical framework, which is
compatible with the GW170817 observational constraints on the
gravitational wave speed. As in previous works, the constraint
that the gravitational wave speed is $c_T^2=1$ in natural units,
results to a constraint differential equation that relates the
coupling function of the scalar field to the Gauss-Bonnet
invariant $\xi(\phi)$ and the scalar potential $V(\phi)$. Adopting
the slow-roll conditions for the scalar field and the Hubble rate,
and in contrast to previous works, by further assuming that
$\kappa \frac{\xi '}{\xi''}\ll 1$, which is motivated by slow-roll
arguments, we succeed in providing much more simpler expressions
for the slow-roll indices and for the tensor and scalar spectral
indices and for the tensor-to-scalar ratio. We exemplify our
refined theoretical framework by using an illustrative example
with a simple power-law scalar coupling function $\xi(\phi)\sim
\phi^{\nu}$ and as we demonstrate the resulting inflationary
phenomenology is compatible with the latest Planck data. Moreover,
this particular model produces a blue-tilted tensor spectral
index, so we discuss in brief the perspective of describing the
NANOGrav result with this model as is indicated in the recent
literature.
\end{abstract}


\maketitle

\section{Introduction}

The last in conjunction with the following decades from an
astronomical point of view belong to the gravitational wave
astronomy branch of physics. Indeed, the LIGO-Virgo collaboration
have already established the new branch of physics
\cite{Harry:2010zz,TheVirgo:2014hva,TheLIGOScientific:2017qsa,Monitor:2017mdv,GBM:2017lvd,LIGOScientific:2019vic},
the gravitational wave astronomy branch, and with the upcoming
LISA \cite{Baker:2019nia,Smith:2019wny} and future space missions
BBO \cite{Crowder:2005nr,Smith:2016jqs} and DECIGO
\cite{Seto:2001qf,Kawamura:2020pcg}, it is expected that our
perception regarding the early Universe and astrophysical
phenomena will be further enlightened. Already, the GW170817 event
\cite{TheLIGOScientific:2017qsa} followed by a kilonova Gamma Ray
Burst \cite{Monitor:2017mdv,GBM:2017lvd} which arrived almost
simultaneously with the gravitational waves, have formed and
changed our perception about modified gravity theories since many
models which generate tensor perturbations with propagation speed
different from that of light's in the vacuum, are excluded from
being phenomenologically viable theories
\cite{Ezquiaga:2017ekz,Baker:2017hug,Creminelli:2017sry,Sakstein:2017xjx}.
Since there is no reason for the gravitons to acquire mass
primordially, and especially during the inflationary era, the
GW170817 event also constrains the primordial gravitational waves
propagation speed $c_T$. Essentially, the tensor perturbations
should also comply with the GW170817 event and thus the
propagation speed of the tensor perturbations must be equal to
unity in natural units, that is $c_T^2\simeq 1$. In view of this
compelling constraint, in a previous work \cite{Odintsov:2020sqy}
we examined the quantitative implications of the constraint
$c_T^2\simeq 1$, for Einstein-Gauss-Bonnet theories
\cite{Hwang:2005hb,Nojiri:2006je,Cognola:2006sp,Nojiri:2005vv,Nojiri:2005jg,Satoh:2007gn,Bamba:2014zoa,Yi:2018gse,Guo:2009uk,Guo:2010jr,Jiang:2013gza,Kanti:2015pda,vandeBruck:2017voa,Kanti:1998jd,Pozdeeva:2020apf,Vernov:2021hxo,Pozdeeva:2021iwc,Fomin:2020hfh,DeLaurentis:2015fea,Chervon:2019sey,Nozari:2017rta,Odintsov:2018zhw,Kawai:1998ab,Yi:2018dhl,vandeBruck:2016xvt,Kleihaus:2019rbg,Bakopoulos:2019tvc,Maeda:2011zn,Bakopoulos:2020dfg,Ai:2020peo,Oikonomou:2020oil,Odintsov:2020xji,Oikonomou:2020sij,Odintsov:2020zkl,Odintsov:2020mkz,Venikoudis:2021irr,Easther:1996yd,Antoniadis:1993jc,Antoniadis:1990uu,Kanti:1995vq,Kanti:1997br}.
As we demonstrated, the most important feature is that the
coupling function of the scalar field to the Gauss-Bonnet
invariant and the scalar potential must be related and cannot be
chosen freely. The resulting theoretical framework yielded a
phenomenologically viable era, and also it yielded quite elegant
expressions for most of the slow-roll indices. More importantly it
offered the possibility of obtaining analytical results assuming
simply the slow-roll conditions, however, the functional forms of
the scalar potential or the scalar coupling function which could
yield viable phenomenologies, were quite involved, and this was
the only drawback of the theoretical framework. In this work we
shall further refine the theoretical framework we developed in
Ref. \cite{Odintsov:2020sqy} by simply imposing one constraint
imposed by the slow-roll approximation. Specifically by assuming
that the scalar coupling function $\xi (\phi)$ satisfies $\kappa
\frac{\xi '}{\xi''}\ll 1$ along with the slow-roll assumption, the
resulting theoretical framework is significantly simplified, and
all the slow-roll indices and the corresponding observational
indices acquire quite simplified final expressions. Moreover, even
simple power-law choices for the scalar coupling function result
to simple expressions for the scalar potential, the slow-roll
indices and the observational indices. In order to illustrate that
the reformed GW170817-compatible Einstein-Gauss-Bonnet framework
we develop in this paper, leads to phenomenologically viable
results, we choose a simple power-law scalar coupling function
$\xi (\phi)$ and we perform a thorough analysis for the resulting
model, eventually confronting the model with the latest (2018)
observational constraints of the Planck collaboration. As we show,
focusing on the spectral index of the primordial scalar curvature
perturbations, and the tensor-to-scalar ratio, the model is
compatible with the Planck 2018 observational data. Moreover, we
discover an interesting feature of this power-law model with
$\xi(\phi)\sim \phi^{\nu}$. Specifically, the model can lead to a
blue tilted tensor spectral index, for large values of the
exponent $\nu$, and this feature is quite interesting with regard
to explaining the NANOGrav results on pulsar timing arrays with a
blue-tilted inflationary theory. We also discuss in brief this
possibility explaining the NANOGrav results with a stochastic
gravitational wave background originating by an inflationary
gravitational wave background corresponding to a blue-tilted
Einstein-Gauss-Bonnet theory.

This paper is organized as follows: In section II we present in
detail the reformed GW170817-compatible Einstein-Gauss-Bonnet
framework. We discuss how the simplified framework is obtained by
assuming that the slow-roll conditions and also the condition
$\kappa \frac{\xi '}{\xi''}\ll 1$ hold true. In section III, we
use a simple power-law model with $x(\phi )\sim \phi^n$ in order
to test the phenomenological viability of the model and confront
it with the latest (2018) Planck observations. We also discuss the
implications of a blue tilt in the tensor spectral index, on the
NANOGrav results. Finally, the conclusions follow at the end of
the paper.

\section{Simplified Einstein-Gauss-Bonnet Inflation Framework}

Consider the Einstein-Gauss-Bonnet action,
\begin{equation}
\label{action} \centering
S=\int{d^4x\sqrt{-g}\left(\frac{R}{2\kappa^2}-\frac{1}{2}\partial_{\mu}\phi\partial^{\mu}\phi-V(\phi)-\frac{1}{2}\xi(\phi)\mathcal{G}\right)}\,
,
\end{equation}
where $R$ stands for the Ricci scalar, $\kappa=\frac{1}{M_p}$
where $M_p$ is the reduced Planck mass, and $\mathcal{G}$ stands
for the Gauss-Bonnet invariant in four dimensions, which is
$\mathcal{G}=R^2-4R_{\alpha\beta}R^{\alpha\beta}+R_{\alpha\beta\gamma\delta}R^{\alpha\beta\gamma\delta}$
with $R_{\alpha\beta}$ and $R_{\alpha\beta\gamma\delta}$ standing
for the Ricci and Riemann tensor respectively. We shall assume
that the background metric is described by a flat
Friedmann-Robertson-Walker (FRW) metric,
\begin{equation}
\label{metric} \centering
ds^2=-dt^2+a(t)^2\sum_{i=1}^{3}{(dx^{i})^2}\, ,
\end{equation}
with $a(t)$ denoting as usual the scale factor. For the FRW
metric, the Gauss-Bonnet invariant takes the form
$\mathcal{G}=24H^2(\dot H+H^2)$, where $H$ is the Hubble rate
which in terms of the scale factor of the FRW metric is defined as
$H=\frac{\dot{a}}{a}$. Also in the following we shall assume that
the scalar field does not have a spatial coordinate dependence.
Upon variation of the gravitational action (\ref{action}) with
respect to the metric and with respect to the scalar field, we
obtain the gravitational equations of motion,
\begin{equation}
\label{motion1} \centering
\frac{3H^2}{\kappa^2}=\frac{1}{2}\dot\phi^2+V+12 \dot\xi H^3\, ,
\end{equation}
\begin{equation}
\label{motion2} \centering \frac{2\dot
H}{\kappa^2}=-\dot\phi^2+4\ddot\xi H^2+8\dot\xi H\dot H-4\dot\xi
H^3\, ,
\end{equation}
\begin{equation}
\label{motion3} \centering \ddot\phi+3H\dot\phi+V'+12 \xi'H^2(\dot
H+H^2)=0\, .
\end{equation}
We shall consider the inflationary era of the
Einstein-Gauss-Bonnet theory, so we assume that the slow-roll
conditions hold true,
\begin{equation}\label{slowrollhubble}
\dot{H}\ll H^2,\,\,\ \frac{\dot\phi^2}{2} \ll V,\,\,\,\ddot\phi\ll
3 H\dot\phi\, .
\end{equation}
Also, in order to render the Einstein-Gauss-Bonnet model
compatible with the GW170817 event, the gravitational wave of the
tensor perturbations, which reads,
\begin{equation}
\label{GW} \centering c_T^2=1-\frac{Q_f}{2Q_t}\, ,
\end{equation}
must be equal to unity in natural units. The functions $Q_f$, $F$
and $Q_b$ defined above are $Q_f=8 (\ddot\xi-H\dot\xi)$,
$Q_t=F+\frac{Q_b}{2}$, $F=\frac{1}{\kappa^2}$ and $Q_b=-8 \dot\xi
H$. Thus in order to have $c_T^2=1$, the condition $Q_f=0$ must
hold true, which results to the differential equation
$\ddot\xi=H\dot\xi$, which constrains the Gauss-Bonnet scalar
coupling function $\xi(\phi)$. This differential equation can be
expressed in terms of the scalar field, using
$\dot\xi=\xi'\dot\phi$ and $\frac{d}{dt}=\dot\phi\frac{d}{d\phi}$,
as follows,
\begin{equation}
\label{constraint1} \centering
\xi''\dot\phi^2+\xi'\ddot\phi=H\xi'\dot\phi\, ,
\end{equation}
and the ``prime'' hereafter will denote differentiation with
respect to the scalar field. By assuming,
\begin{equation}\label{firstslowroll}
 \xi'\ddot\phi \ll\xi''\dot\phi^2\, ,
\end{equation}
which is strongly motivated by the slow-roll conditions of the
scalar field, the constraint of Eq. (\ref{constraint1}) becomes,
\begin{equation}
\label{constraint} \centering
\dot{\phi}\simeq\frac{H\xi'}{\xi''}\, .
\end{equation}
Upon combining Eqs. (\ref{motion3}) and (\ref{constraint}) we
obtain,
\begin{equation}
\label{motion4} \centering \frac{\xi'}{\xi''}\simeq-\frac{1}{3
H^2}\left(V'+12 \xi'H^4\right)\, .
\end{equation}
So far this framework was used in previous works
\cite{Odintsov:2020sqy,Oikonomou:2020oil,Odintsov:2020xji,Oikonomou:2020sij,Odintsov:2020zkl,Odintsov:2020mkz},
at this point we shall differentiate from the previous studies
based on the fact the first slow-roll index for this system
depends on the ratio $\xi'/\xi''$. Specifically, we shall consider
theories for which,
\begin{equation}\label{mainnewassumption}
\kappa \frac{\xi '}{\xi''}\ll 1\, ,
\end{equation}
and in addition the following extra condition,
\begin{equation}\label{scalarfieldslowrollextra}
12 \dot\xi H^3=12 \frac{\xi'^2H^4}{\xi''}\ll V\, ,
\end{equation}
which is strongly related to the constraint
(\ref{mainnewassumption}). Upon combining Eqs.
(\ref{slowrollhubble}), (\ref{constraint}) and
(\ref{scalarfieldslowrollextra}), we can rewrite the equations of
motion in the following quite simplified form,
\begin{equation}
\label{motion5} \centering H^2\simeq\frac{\kappa^2V}{3}\, ,
\end{equation}
\begin{equation}
\label{motion6} \centering \dot H\simeq-\frac{1}{2}\kappa^2
\dot\phi^2\, ,
\end{equation}
\begin{equation}
\label{motion8} \centering \dot\phi\simeq\frac{H\xi'}{\xi''}\, .
\end{equation}
In view of Eq. (\ref{motion5}), the condition
(\ref{scalarfieldslowrollextra}) takes the simpler form,
\begin{equation}\label{mainconstraint2}
 \frac{4\kappa^4\xi'^2V}{3\xi''}\ll 1\, ,
\end{equation}
which shall be extensively used in the following analysis. More
importantly, the differential equation (\ref{motion4}), which
essentially connects the scalar coupling function $\xi(\phi)$ with
the scalar potential, becomes,
\begin{equation}
\label{maindiffeqnnew} \centering
\frac{V'}{V^2}+\frac{4\kappa^4}{3}\xi'\simeq 0\, ,
\end{equation}
which must be obeyed by both the scalar coupling function $\xi
(\phi)$ and the scalar potential, and essentially it is very
important for the analysis that follows.

Let us proceed to demonstrate how the whole Einstein-Gauss-Bonnet
inflationary framework is simplified, so let us start by recalling
the definition of the slow-roll indices \cite{Hwang:2005hb},
\begin{align}\label{slowrollbasic}
& \epsilon_1=-\frac{\dot
H}{H^2},\,\,\,\epsilon_2=\frac{\ddot\phi}{H\dot\phi}, \,\,\,
\epsilon_3=\frac{\dot F}{2HF}, \, \epsilon_4=\frac{\dot E}{2HE},
\\ \notag &
\epsilon_5=\frac{\dot F+Q_a}{2HQ_t},\,\,\, \epsilon_6=\frac{\dot
Q_t}{2HQ_t}\, ,
\end{align}
with $F=\frac{1}{\kappa^2}$, and also $E$ is defined as follows,
\begin{equation}\label{functionE}
E=\frac{F}{\dot\phi^2}\left( \dot{\phi}^2+3\left(\frac{(\dot
F+Q_a)^2}{2Q_t}\right)\right)\, ,
\end{equation}
and in addition $Q_a$, $Q_t$, $Q_b$ and $Q_e$ are defined as
follows \cite{Hwang:2005hb},
\begin{align}\label{qis}
& Q_a=-4 \dot\xi H^2,\,\,\,Q_b=-8 \dot\xi H,\,\,\,
Q_t=F+\frac{Q_b}{2},\,\,\, Q_e=-16 \dot{\xi}\dot{H}\, .
\end{align}
By using the simplified equations of motion, and Eqs.
(\ref{mainnewassumption}) and (\ref{mainconstraint2}), the
slow-roll indices take the following simplified form (after some
algebra),
\begin{equation}
\label{index1} \centering \epsilon_1\simeq\frac{\kappa^2
}{2}\left(\frac{\xi'}{\xi''}\right)^2\, ,
\end{equation}
\begin{equation}
\label{index2} \centering
\epsilon_2\simeq1-\epsilon_1-\frac{\xi'\xi'''}{\xi''^2}\, ,
\end{equation}
\begin{equation}
\label{index3} \centering \epsilon_3=0\, ,
\end{equation}
\begin{equation}
\label{index4} \centering
\epsilon_4\simeq\frac{\xi'}{2\xi''}\frac{\mathcal{E}'}{\mathcal{E}}\,
,
\end{equation}
\begin{equation}
\label{index5} \centering
\epsilon_5\simeq-\frac{\epsilon_1}{\lambda}\, ,
\end{equation}
\begin{equation}
\label{index6} \centering \epsilon_6\simeq
\epsilon_5(1-\epsilon_1)\, ,
\end{equation}
where we defined the functions of $\phi$,
$\mathcal{E}=\mathcal{E}(\phi)$ and $\lambda=\lambda(\phi)$ as
follows,
\begin{equation}\label{functionE}
\mathcal{E}(\phi)=\frac{1}{\kappa^2}\left(
1+72\frac{\epsilon_1^2}{\lambda^2} \right),\,\, \,
\lambda(\phi)=\frac{3}{4\xi''\kappa^2 V}\, .
\end{equation}
Having derived the simple expressions for the slow-roll indices,
we can directly find the observational indices of inflation for
the model at hand. We shall be interested in the spectral index of
the primordial scalar perturbation $n_{\mathcal{S}}$, the spectral
index of the primordial tensor perturbations $n_{\mathcal{T}}$ and
the tensor-to-scalar ratio $r$, which in terms of the slow-roll
indices have the following form,
 \cite{Hwang:2005hb},
\begin{equation}
\label{spectralindex} \centering
n_{\mathcal{S}}=1-4\epsilon_1-2\epsilon_2-2\epsilon_4\, ,
\end{equation}
\begin{equation}\label{tensorspectralindex}
n_T=-2\left( \epsilon_1+\epsilon_6 \right)\, ,
\end{equation}
\begin{equation}\label{tensortoscalar}
r=16\left|\left(\frac{\kappa^2Q_e}{4H}-\epsilon_1\right)\frac{2c_A^3}{2+\kappa^2Q_b}\right|\,
,
\end{equation}
where $c_A$ is the sound speed defined as follows,
\begin{equation}
\label{sound} \centering c_A^2=1+\frac{Q_aQ_e}{3Q_a^2+
\dot\phi^2(\frac{2}{\kappa^2}+Q_b)}\, .
\end{equation}
Notice the complicated form of the tensor-to-scalar ratio which is
not as simple as the one corresponding to the single scalar case.
In our case, since the theory is a Einstein-Gauss-Bonnet theory,
the tensor-to-scalar ratio depends on the terms $Q_a$, $Q_b$ and
$Q_e$, explicitly and via the sound wave speed $c_A$. However, by
using the conditions (\ref{mainconstraint2}) and
(\ref{mainnewassumption}) in conjunction with the slow-roll
conditions, after some algebra it can be shown that the final
expression of the tensor-to-scalar ratio is,
\begin{equation}\label{tensortoscalarratiofinal}
r\simeq 16\epsilon_1\, ,
\end{equation}
which is functionally identical to the single scalar field case,
with, recall, $\epsilon_1$ given in Eq. (\ref{index1}). Also by
using the functional form of the simplified slow-roll index
$\epsilon_6$ from Eqs. (\ref{index5}) and (\ref{index6}), one can
easily obtain the final expression for the spectral index of the
tensor perturbations,
\begin{equation}\label{tensorspectralindexfinal}
n_{\mathcal{T}}\simeq -2\epsilon_1\left ( 1-\frac{1}{\lambda
}+\frac{\epsilon_1}{\lambda}\right)\, ,
\end{equation}
where the function $\lambda$ is defined in Eq. (\ref{functionE}).
A direct comparison of the final expressions for the slow-roll
indices (\ref{index1})-(\ref{index6}), and of the tensor spectral
index (\ref{tensorspectralindexfinal}) and the tensor-to-scalar
ratio (\ref{tensortoscalarratiofinal}), with the expressions
obtained in Ref. \cite{Odintsov:2020sqy}, can show the simplicity
of the expressions obtained in the present work. Finally, the
$e$-foldings number has the form,
\begin{equation}
\label{efolds} \centering
N=\int_{t_i}^{t_f}{Hdt}=\int_{\phi_i}^{\phi_f}\frac{H}{\dot{\phi}}d\phi=\int_{\phi_i}^{\phi_f}{\frac{\xi''}{\xi'}d\phi}\,
,
\end{equation}
which is the same compared to the one presented in Ref.
\cite{Odintsov:2020sqy}, with $\phi_f$ being the value of the
scalar field at the end of the inflationary era, and $\phi_i$ is
the value of the scalar field at the beginning of the inflationary
era, precisely at the first horizon crossing time instance.
Essentially, the slow-roll indices (\ref{index1})-(\ref{index6}),
the tensor spectral index (\ref{tensorspectralindexfinal}) and the
tensor-to-scalar ratio (\ref{tensortoscalarratiofinal}) are the
main results of this work. We shall use this simplified framework
to examine whether a viable phenomenology can be obtained. As we
will see in the next section, by using relatively simple
expressions for the potential and the scalar coupling function,
can lead to a viable inflationary era.

\section{Confrontation with Observations: A Power-law Einstein-Gauss-Bonnet Model with Blue-tilted Tensor Spectral Index}

\begin{figure}
\centering
\includegraphics[width=18pc]{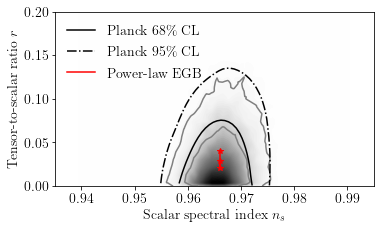}
\caption{Marginalized curves of the Planck 2018 data and the model
confronted with the Planck 2018 data (red
curve).}\label{plotplanck}
\end{figure}

In this section we shall demonstrate that the simplified
Einstein-Gauss-Bonnet theoretical framework we introduced in the
previous section, can produce a phenomenologically viable
inflationary era. We shall consider the simplest case by choosing
the scalar coupling function to have a power-law dependence with
respect to the scalar field, and we shall examine in detail the
inflationary predictions of the model. As we will show, the model
apart from being viable, it also produces a blue-tilted tensor
spectral index, and we discuss in brief the possibility of
explaining the NANOGrav observations
\cite{Arzoumanian:2020vkk,NANOGrav:2020spf} with this blue tilted
model. Also, as we show, the resulting blue-tilted tensor spectrum
is also compatible with the LIGO-Virgo constraints
\cite{Harry:2010zz,TheVirgo:2014hva,TheLIGOScientific:2017qsa,Monitor:2017mdv,GBM:2017lvd}.
With regard to the inflationary era observational quantities, we
shall be interested in the spectral index of the primordial scalar
curvature perturbations $n_{\mathcal{S}}$, the tensor-to-scalar
ratio $r$, and the tensor spectral index which for the theoretical
framework at hand are given in Eqs. (\ref{spectralindex}),
(\ref{tensortoscalarratiofinal}) and
(\ref{tensorspectralindexfinal}) respectively. Recall that the
scalar spectral index and the tensor-to-scalar ratio are
constrained by the 2018 Planck data \cite{Akrami:2018odb} as
follows,
\begin{equation}\label{planck2018}
\centering n_{\mathcal{S}}=0.9649\pm0.0042,\,\,\, r<0.064\, .
\end{equation}
Now to proceed with the power-law model, let us assume that the
scalar coupling function has the following power-law form,
\begin{equation}
\label{modelA} \xi(\phi)=\beta  (\kappa  \phi )^{\nu }\, ,
\end{equation}
where $\beta$ is a dimensionless parameter, while recall that
$\kappa=1/M_p$. By substituting the scalar coupling (\ref{modelA})
in Eq. (\ref{maindiffeqnnew}) by solving the differential
equation, we can obtain the scalar potential, which reads,
\begin{equation}
\label{potA} \centering V(\phi)=\frac{3}{4 \beta  \kappa ^{\nu +4}
\phi ^{\nu }+3 \gamma  \kappa ^4} \, ,
\end{equation}
where $\gamma$ is a dimensionless integration constant. From both
Eqs. (\ref{modelA}) and (\ref{potA}) it is apparent that both the
scalar coupling function and the scalar potential have
particularly simple form, in contrast to the more involved cases
studied in Ref. \cite{Odintsov:2020sqy}. In addition, for these
choices of the scalar functions, the slow-roll indices
(\ref{index1})-(\ref{index6}) acquire particularly simple
functional forms, which we quote here,
\begin{equation}
\label{index1A} \centering \epsilon_1\simeq \frac{\kappa ^2 \phi
^2}{2 (\nu -1)^2} \, ,
\end{equation}
\begin{equation}
\label{index2A} \centering \epsilon_2\simeq -\frac{\kappa ^2 \phi
^2-2 \nu +2}{2 (\nu -1)^2}\, ,
\end{equation}
\begin{equation}
\label{index3A} \centering \epsilon_3=0\, ,
\end{equation}
\begin{equation}
\label{index4A} \centering \epsilon_4\simeq \frac{\phi
\left(\kappa \, (2 \nu -4) \alpha (\phi ) \zeta (\phi )-8 \beta
\nu \zeta (\phi ) \kappa ^{\nu +5} \phi ^{\nu }\right)}{2 \kappa
(\nu -1) \phi  \alpha (\phi ) (\zeta (\phi )+1)} \, ,
\end{equation}
\begin{equation}
\label{index5A} \centering \epsilon_5\simeq -\frac{2 \beta \,
\kappa^4\, \nu (\kappa  \phi )^{\nu }}{(\nu -1)  \alpha (\phi )}
\, ,
\end{equation}
\begin{equation}
\label{index6A} \centering \epsilon_6\simeq -\frac{\beta \,
\kappa^4\, \nu (\kappa  \phi )^{\nu } \left(-\kappa ^2 \phi ^2+2
\nu ^2-4 \nu +2\right)}{(\nu -1)^3 \alpha (\phi )} \, ,
\end{equation}
which are quite simple functionally, compared to Ref.
\cite{Odintsov:2020sqy}. In the slow-roll index $\epsilon_4$, the
functions $\zeta (\phi)$ and $\alpha (\phi)$ are defined as
follows, $\zeta (\phi)=\frac{288 \beta ^2 \kappa ^{12} \nu ^2 \phi
^4 (\kappa \phi )^{2 \nu -4}}{(\nu -1)^2 \alpha (\phi )^2}$ and
$\alpha (\phi)=4 \beta \kappa ^{\nu +4} \phi ^{\nu }+3 \gamma
\kappa ^4$. Upon solving $\epsilon_1\simeq \mathcal{O}(1)$ we get
the value of the scalar field at the end of the inflationary era
which reads $\phi_f\simeq \frac{\sqrt{2} (\nu-1)}{\kappa }$, and
also the value of the scalar field $\phi_i$ at the first horizon
crossing during the early stages of the inflationary era, can be
obtained by using the definition of the $e$-foldings number in Eq.
(\ref{efolds}). Thus upon solving Eq. (\ref{efolds}) with respect
to $\phi_i$ we obtain $\phi_i=\frac{\sqrt{2} (\nu-1)
e^{-\frac{N}{\nu -1}}}{\kappa }$. Accordingly, by evaluating the
scalar spectral index at the first horizon crossing, it reads,
\begin{align}\label{spectralpowerlawmodel}
& n_{\mathcal{S}}\simeq -1+\frac{2 (\nu -2)}{\nu -1}-2\,
e^{-\frac{2 N}{\nu -1}} + (\nu -1)^{2 \nu -3} \nu ^2 \times \\
\notag &  \frac{9\, \beta ^2 \,2^{\nu +6} \,\left(\beta \,
2^{\frac{\nu }{2}+3}\, (\nu -1)^{\nu +1}+3 \gamma  (\nu -2)
e^{\frac{\nu N}{\nu -1}}\right)}{\left(\beta \, 2^{\frac{\nu
}{2}+2}\, (\nu -1)^{\nu }+3 \gamma  e^{\frac{\nu  N}{\nu
-1}}\right)^3} \, ,
\end{align}
while the tensor-to-scalar ratio has the simple form,
\begin{equation}\label{tensortoscalarfinalmodelpowerlaw}
r\simeq 16\, e^{-\frac{2 N}{\nu -1}}\, .
\end{equation}
Finally, the tensor-spectral index calculated at the first horizon
crossing, has the following form,
\begin{align}\label{tensorspectralindexpowerlawmodel}
& n_{\mathcal{T}}\simeq \frac{12 \gamma  e^{\frac{(\nu -4) N}{\nu
-1}}}{\beta \, 2^{\frac{\nu }{2}+2} (\nu -1)^{\nu }+3 \gamma
e^{\frac{\nu  N}{\nu -1}}} \\ \notag & -\frac{\beta \,
2^{\frac{\nu }{2}+3} (\nu -1)^{\nu -1} e^{-\frac{4 N}{\nu -1}}
\left(-3 \nu +(\nu -1) \nu  e^{\frac{4 N}{\nu
-1}}+2\right)}{\beta\, 2^{\frac{\nu }{2}+2} (\nu -1)^{\nu }+3
\gamma  e^{\frac{\nu N}{\nu -1}}} \, .
\end{align}
One can assign several values to the free dimensionless parameters
$\beta$, $\gamma$ and $\nu$ and the model can be confronted with
the observational data. Assuming $N\simeq 60$, there is a large
range of free parameters which can generate a phenomenologically
viable inflationary era. We shall consider four sets of parameter
values for the parameters $\nu$ and $\gamma$, which are the
following,
\begin{equation}\label{set1}
\mathrm{set}\,\,1\,\,\,\,\nu=20,\,\,\,\gamma=5768\, ,
\end{equation}
\begin{equation}\label{set2}
\mathrm{set}\,\,2\,\,\,\,\nu=19,\,\,\,\gamma=57\, ,
\end{equation}
\begin{equation}\label{set3}
\mathrm{set}\,\,3\,\,\,\,\nu=19,\,\,\,\gamma=5\, ,
\end{equation}
\begin{equation}\label{set4}
\mathrm{set}\,\,4\,\,\,\,\nu=21,\,\,\,\gamma=7.7\times 10^8\, ,
\end{equation}
Let us consider in some detail the parameter set 1. The spectral
index for $\nu$ and $\gamma$ taking the values specified in set 1
(\ref{set1}) takes the following form,
\begin{equation}\label{spectralindexset1}
n_{\mathcal{S}}=\frac{7.90773\times 10^6 \beta ^3}{(57.3201 \beta
+17305.5)^3}+\frac{124162. \beta ^2}{(57.3201 \beta
+17305.5)^2}+0.891122\, ,
\end{equation}
so if we demand that the spectral index takes the value
$n_s=0.966$ there are three values of the parameter $\beta$ that
yield this result, namely, $\beta=-13.191$, $\beta=-142.825$ and
$\beta=13.722$. After we performed our analysis, we ended up to
the result that the tensor spectral index is red-tilted for
negative values of the parameter $\beta$, while it is blue tilted
for positive values of the parameter $\beta$. This result holds
true for all the sets of parameter values, that is for set 2 and
set 3. Let us present one indicative case of red-tilted tensor
spectrum, for example by choosing $\beta=-13.191$ we get,
$n_\mathcal{S}=0.966$, $n_{\mathcal{T}}=-0.0516207$, $r=0.0289206$
which are all compatible with the Planck 2018 constraints
(\ref{planck2018}). Now in order to validate that all the
approximations we made in the previous section hold true, later on
in this section we discuss this issue in detail and in Table
\ref{table1} we present the order of magnitude of the
approximations we made in the previous section in order to derive
the formulas, calculated at the horizon crossing time instance.
For the same set of parameter values, namely set 1, for
$\beta=13.722$ we obtain a blue tilted spectrum, and specifically
the observational indices read, $n_\mathcal{S}=0.966$,
$n_{\mathcal{T}}=0.0420649$, $r=0.0289206$. Notice that the
tensor-to-scalar ratio is the same, which is expected since it
solely depends on the parameters $\nu$ and $N$. We shall further
discuss this issue in the end of this section. Now let us proceed
to the rest of the parameter sets, starting with set 2, and
hereafter we shall focus solely on the blue-tilted spectrum cases.
So for set 2 (\ref{set2}) if we choose $\beta=12.5016$ we get,
$n_\mathcal{S}=0.966$, $n_{\mathcal{T}}=0.0446284$, $r=0.0203621$
which are again compatible with the Planck data. Proceeding with
set 3 (\ref{set3}), for $\beta=1.096$ $n_\mathcal{S}=0.966$,
$n_{\mathcal{T}}=0.0446047$, and the same tensor-to-scalar ratio,
thus viability is obtained for a wide range of parameter values.
Proceeding with set 4 (\ref{set4}), for $\beta=19312$
$n_\mathcal{S}=0.966$, $n_{\mathcal{T}}=0.0394508$, and
$r=0.03966$, thus viability is obtained for a wide range of
parameter values. In order to see visually how well do the
aforementioned set of values fit the latest Planck data, in Fig.
\ref{plotplanck} we present the marginalized curves of Planck data
and the red curve correspond to the above three sets of values for
$N=60$.
Now let us consider the approximations issue, which is definitely
an important issue to discuss in some detail. In our analysis
above all the approximations used in this work, are satisfied for
the values of the free parameters that guarantee the inflationary
viability of the model, as the reader can convince himself.
However in order to make the article more self-contained, let us
show here in detail that all the approximations we assumed in this
paper, are satisfied during the first horizon crossing, and also
we shall calculate the values of all the slow-roll indices during
the first horizon crossing  in order to show that the slow-roll
conditions hold true during the first horizon crossing. Before
getting into the details of our analysis, let us discuss an
important issue, having to do with the slow-roll era of inflation.
All the slow-roll approximations and all the approximations made
in this article, are required to hold true at the first horizon
crossing. Basically the first horizon crossing is relevant for the
present day observations, but as the inflationary era proceeds,
the slow-roll approximation and all the rest of the approximations
will be violated when the slow-roll approximation is violated.
This instance is quantitatively when $\epsilon_1\sim
\mathcal{O}(1)$. Thus it is natural near the end of inflation,
when approximately $\epsilon_1\sim \mathcal{O}(1)$, hence in our
analysis we shall concentrate only on values of the approximations
at the first horizon crossing, which is relevant for the present
day observations, and all the observational indices are calculated
at the first horizon crossing, for $\phi=\phi_i=\frac{\sqrt{2}
(\nu-1) e^{-\frac{N}{\nu -1}}}{\kappa }$. Let us recall that the
modes that exited the Hubble horizon when $\phi=\phi_i$ are the
modes that are relevant for present day observations. These are
the modes that correspond to $k\sim 0.05$Mp$^{-1}$, hence
basically these modes re-enter the horizon when the CMB was
generated, so at $z\sim 1100$. Basically these are highly
short-length ancient primordial modes and the CMB gives exact
details on these modes. This is how the CMB can constrain
inflation itself, through these modes. Let us comment though that
unfortunately, to date no tensor modes have ever been observed.

Let us begin with the case set 1, so for $\beta=13.722$, we have
$\phi_i=1.14238 M_p$ and $\phi_f=26.8701 M_p$, and also the
slow-roll indices at the first horizon crossing, thus when
$\phi=\phi_i$, take the values
$(\epsilon_1,\epsilon_2,\epsilon_4,\epsilon_5,\epsilon_6)=(0.00180754,
0.050824, 0.0327524, -0.0228814, -0.02284)$. Let us note here that
the values of the scalar field are of the Planck scale, and in the
end of this section we shall briefly discuss if these values lead
to trans-Planckian inconsistencies, the same applies for the rest
of the parameter values sets. Regarding set 2 we have
$\phi_i=0.908112 M_p$ and $\phi_f=25.4558 M_p$, and also the
slow-roll indices at the first horizon crossing, take the values
$(\epsilon_1,\epsilon_2,\epsilon_4,\epsilon_5,\epsilon_6)=(0.00127263,
0.0542829, 0.0346397, -0.0236169, -0.0235869)$. Regarding the set
3, we have $\phi_i=0.908112 M_p$ and $\phi_f=25.4558 M_p$,
basically these are the same as in set 2 due to the fact that
$\nu$ is the same. Also the slow-roll indices at the first horizon
crossing, take the values
$(\epsilon_1,\epsilon_2,\epsilon_4,\epsilon_5,\epsilon_6)=(0.00127263,
0.0542829, 0.034607, -0.023605, -0.023575)$. Finally for set 4,
for $\beta=19312$, we have $\phi_i=1.40819 M_p$ and
$\phi_f=28.2843 M_p$, and also the slow-roll indices at the first
horizon crossing, thus when $\phi=\phi_i$, take the values
$(\epsilon_1,\epsilon_2,\epsilon_4,\epsilon_5,\epsilon_6)=(0.00247875,
0.0475212, 0.0311898, -0.0222593, -0.0222041)$. Therefore the
slow-roll conditions hold well true at the first horizon crossing.
Let us now consider the quantities entering the approximations
quoted in Eqs. (\ref{slowrollhubble}), (\ref{firstslowroll}),
(\ref{mainnewassumption}), (\ref{scalarfieldslowrollextra}) and
(\ref{mainconstraint2}), which we also quote here for convenience,
\begin{align}
& \frac{\dot\phi^2}{2} \ll V\, , \,\,\,\kappa \frac{\xi
'}{\xi''}\ll 1,\, ,\,\,\, \frac{4\kappa^4\xi'^2V}{3\xi''}\ll 1 \,
.
\end{align}
In Table \ref{table1} we present all the results of our analysis
regarding the approximations. As it can be seen, all the
approximations we made hold well true at the first horizon
crossing. Some values are identical but not accidentally, it is
due to the functional form of the corresponding approximation, for
example $\kappa \frac{\xi '}{\xi''}$ at the first horizon
crossing.
\begin{table}[h!]
  \begin{center}
    \caption{\emph{\textbf{Values of the Various Approximations at First Horizon Crossing}}}
    \label{table1}
    \begin{tabular}{|r|r|r|r|r|r|}
     \hline
      \textbf{Physical Quantities}   & \textbf{Set 1} $\beta=-13.191$ & \textbf{Set 1} $\beta=13.722$ & \textbf{Set 2} $\beta=12.5016$  & \textbf{Set 3}
      $\beta=1.096$& \textbf{Set 4}
      $\beta=19312$
      \\  \hline
      $\frac{V(\phi)}{\frac{\dot\phi^2}{2}}$ & 3027.47 & 1659.72 & 2357.32
      & 2357.32 & 1210.29
 \\  \hline
       $\kappa \frac{\xi '}{\xi''}$ & 0.0100209 & 0.0100209 & 0.00840844
       &0.00840844 & 0.0117349
      \\  \hline
      $\frac{4\kappa^4\xi'^2V}{3\xi''}$& -0.0480926 & 0.0457627 & 0.0472338
      &0.04721 & 0.0445186
      \\  \hline
    \end{tabular}
  \end{center}
\end{table}
A notable feature of this simple Einstein-Gauss-Bonnet model is
that it yields a blue-tilted tensor spectral index. Although the
Planck data do not constrain the tensor spectral index, for
red-tilted values, blue-tilted values of the tensor spectral index
are constrained by the LIGO-Virgo observational data
\cite{TheLIGOScientific:2017qsa,Monitor:2017mdv,GBM:2017lvd}. In
the next subsection we discuss in brief the phenomenological
outcomes of this model in view of the LIGO-Virgo data, and also
the possibility of explaining the NANOGrav results
\cite{Arzoumanian:2020vkk,NANOGrav:2020spf}. Moreover, let us note
that the blue-tilted tensor spectral index is obtained for several
values of the exponent $\nu$ in Eq. (\ref{modelA}) and basically
the blue tilt is  determined partially by the values of the
parameter $\nu$, but mainly from the values of $\beta$.
Specifically the negative values of $\beta$ yield a red-tilted
tensor spectral index, while the positive values of $\beta$ yield
a blue-tilt for the tensor spectral index. The parameter $\nu$
indirectly affects the presence of a blue spectrum since for
$\nu\leq 3$, only positive values of $\beta$ are obtained, thus a
blue-tilted tensor spectral index is obtained.

Also we need to note that for $\nu\geq 23$ the model become
incompatible with the Planck data for $N=60$, since the
tensor-to-scalar ratio becomes quite larger than the Planck
constraints. Finally let us discuss the issue of having large
values for the scalar field during inflation and more importantly
the fact that the scalar field value increases during inflation.
The latter seems to be a model dependent issue, being related to
the combined presence of the potential and of the Gauss-Bonnet
coupling. There is no general rule to our knowledge that forbids
this general behavior. It might be possible that the scalar field
slow-rolls the potential and at the value $\phi_f$ which is larger
than $\phi_i$, the slow-roll ends and the scalar field starts to
oscillate to produce reheating. However the combined presence of
the scalar potential and the Gauss-Bonnet coupling makes the
interpretation difficult. The situation is quite perplexed
compared with the simple scalar field rolling down its potential
to reach the minimum. Another highly related question is whether
the slow-roll approximated theory we developed above is a stable
limit of the original theory, or whether it leads to some stable
de Sitter vacuum. This is hard to tell in the context of our
formalism and the only rigid answer can be given once the complete
phase space of the model is studied with the constraint $c_T^2=1$.
We aim to perform this analysis in a separate work focused exactly
on this, but a first comment is that during inflation, it would be
desirable to have an exact unstable de Sitter attractor or at
least some unstable quasi-de Sitter attractor, as was shown for
the $R^2$ model in $f(R)$ gravity in Ref. \cite{Odintsov:2017tbc}.
With regard to the large values of the scalar field of the order
$\mathcal{O}(20\,M_p)$, during inflation it is expected that the
scalar field takes values of the order of Planck mass but such
large values are notable, and should be considered in view of the
trans-Planckian aspects of the theory. This study however extends
our knowledge and aims of this introductory paper, and we hope to
address in the future.

Before closing let us consider another important issue that we did
not consider previously, having to do with the amplitude of the
scalar perturbations $\mathcal{P}_{\zeta}(k_*)$,
\begin{equation}\label{definitionofscalaramplitude}
\mathcal{P}_{\zeta}(k_*)=\frac{k_*^3}{2\pi^2}P_{\zeta}(k_*)
\end{equation}
evaluated at the pivot scale $k_*=0.05$Mpc$^{-1}$, which is
relevant for CMB observations. This is constrained by Planck to
take values in $\mathcal{P}_{\zeta}(k_*)=2.196^{+0.051}_{-0.06}$.
The reason we chose four sets of parameter values is not
accidental. As we will show now, although sets 1-3 provide a
viable phenomenology, they produce too small amplitude for the
scalar perturbations, and only set 4 provides a viable amplitude
for the scalar perturbations. However, we quote this analysis
last, because this amplitude is more or less $\Lambda$CDM
connected. Recall that the scalar amplitude for the scalar
perturbations $\mathcal{P}_{\zeta}(k)$ is related to the two point
function for the curvature perturbations $\zeta (k)$ via
$P_{\zeta}(k)$ appearing in Eq.
(\ref{definitionofscalaramplitude}) as follows,
\begin{equation}\label{twopointfunctionforzeta}
\langle\zeta(k)\zeta (k')\rangle=(2\pi)^3 \delta^3 (k-k')
P_{\zeta}(k)\, .
\end{equation}
Now let us see the predictions for $\mathcal{P}_{\zeta}(k_*)$ for
the four different sets of parameter values, evaluated at the
pivot scale $k_*$. The formula for $\mathcal{P}_{\zeta}(k_*)$ for
the Einstein-Gauss-Bonnet model at hand can be found in Ref.
\cite{Hwang:2005hb} and in the slow-roll approximation it is,
\begin{equation}\label{powerspectrumscalaramplitude}
\mathcal{P}_{\zeta}(k)=\left(\frac{k \left((-2
\epsilon_1-\epsilon_2-\epsilon_4) \left(0.57\, +\log \left(\left|
\frac{1}{1-\epsilon_1}\right| \right)-2+\log
(2)\right)-\epsilon_1+1\right)}{(2 \pi ) \left(z
c_A^{\frac{4-n_{\mathcal{S}}}{2}}\right)}\right)^2\, ,
\end{equation}
where $z=\frac{(\dot{\phi} k) \sqrt{\frac{E(\phi
)}{\frac{1}{\kappa ^2}}}}{H^2 (\epsilon_5+1)}$, and we used $k=aH$
at the horizon crossing, and also the conformal time at horizon
crossing is $\eta=-\frac{1}{aH}\frac{1}{-\epsilon_1+1}$. We shall
evaluate it at the first horizon crossing, with $k_*=aH$. For set
1, and $\beta=13.722$ we obtain
$\mathcal{P}_{\zeta}(k_*)=0.000402806$, for set 2 and
$\beta=12.051$ we get $\mathcal{P}_{\zeta}(k_*)=0.0579883$, for
set 3 and $\beta=1.096$ we obtain
$\mathcal{P}_{\zeta}(k_*)=0.661094$, and for set 4 and
$\beta=19312$ we get $\mathcal{P}_{\zeta}(k_*)=2.19673\times
10^{-9}$. Obviously only set 4 produces a viable scalar
perturbations amplitude, thus it seems that the model's viability
is crucially affected by the values of $\beta$ and $\gamma$, while
$\nu$ does not crucially affect the results. Hence, if one also
requires the amplitude of the scalar perturbations to be
compatible with the Planck data, one must require large values of
the parameter $\gamma$. In fact, when $\gamma$ takes smaller
values, the scalar amplitude of curvature perturbations deviates
more and more from the Planck constraints. Furthermore let us
comment that the approximations we made and eventually the
slow-roll conditions seize to hold true at the end of the
inflationary era, as it is expected.

\subsection{Blue Tilt and Discussion in View of the NANOGrav Results}

\begin{figure}
\centering
\includegraphics[width=18pc]{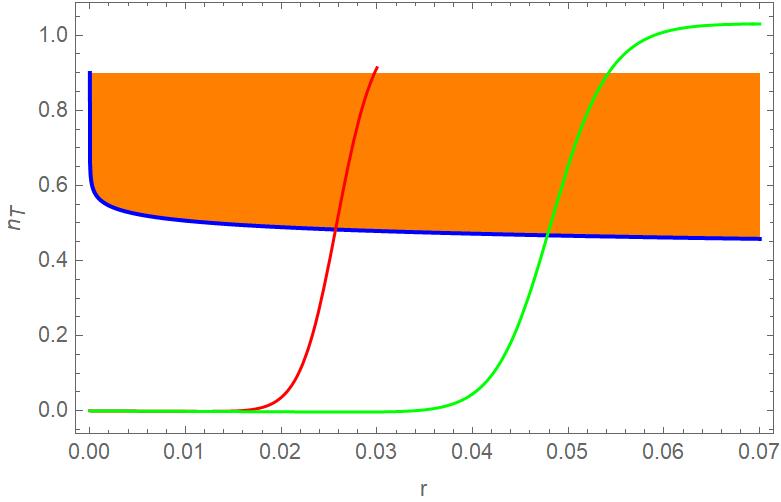}
\caption{The tensor spectral index constraints from the LIGO-Virgo
collaboration (blue curve). Values in the orange shaded area are
excluded by the LIGO-Virgo data and the blue curve indicates the
maximum upper limit of the allowed values for the tensor spectral
index. The green and red curves correspond to values of the tensor
spectral index for the power-law Einstein-Gauss-Bonnet model of
Eqs. (\ref{modelA}) and (\ref{potA}).}\label{plot1}
\end{figure}

One of the most astonishing future observations will, hopefully,
be the detection of a stochastic primordial gravitational wave
background. To date, the frequencies tested already by the
LIGO-Virgo collaboration \cite{LIGOScientific:2019vic} put strong
constraints on the stochastic gravitational wave spectrum
\begin{equation}\label{spectralequatio}
\Omega_{\mathcal{G}\mathcal{W}}(k)=\frac{1}{\rho_c}\frac{d
\rho_{\mathcal{G}\mathcal{W}}}{ d \log k}\, , \notag
\end{equation}
for scales $k_{LV}$ in the range $(1.3-5.5)\times
10^{16}$Mpc$^{-1}$, and specifically,
\begin{equation}\label{constraintligovirgo}
\Omega_{\mathcal{G}\mathcal{W}}(k)\leq 1.7\times 10^{-7}\, ,
\end{equation}
at $95\%$CL. Essentially, these scales $k_{LV}$ correspond to
quite large wavelengths originating back to the first stages of
the radiation era, and possibly during the speculated reheating
era. The frequencies of the primordial modes $k_{LV}$ are
basically $20-84$Hz, the lowest range of the LIGO-Virgo
sensitivities, and some serious constraints can be imposed on the
tensor spectral index. The Cosmic Microwave Background radiation
scales correspond to much lower frequencies, and much large
wavelengths that entered the horizon at $k_*=0.05\,$Mpc$^{-1}$,
and only weakly constrain the tensor spectral index, but the
LIGO-Virgo for a power-law tensor spectrum, for
$\mathcal{P}_T(k_*)=r A_S$, where $A_S$ is the amplitude of the
scalar perturbations at $k_*$, constrain the tensor spectral index
as follows \cite{Giare:2020vss},
\begin{equation}\label{spectrumGWconstraintontensor}
n_{\mathcal{T}}<\frac{\ln \left (
\frac{2z_{eq}\Omega_{\mathcal{G}\mathcal{W}}(k_{LV})}{rA_S}\right)}{\ln
\left(\frac{k_{LV}}{k_*} \right)}\leq 0.5\, ,
\end{equation}
where $z_{eq}$ is the redshift at the matter-radiation equality,
and $A_S=2.1\times 10^{-9}$ by the latest Planck data. Upon taking
the LIGO-Virgo constraint into account on the spectrum, in Fig.
\ref{plot1} we plot the dependence of the constraint
(\ref{spectrumGWconstraintontensor}) from the tensor-to-scalar
ratio, and in the shaded area belong the values of
$n_{\mathcal{T}}$ which are excluded by the latest LIGO-Virgo
observational constraints. The lower limit of the constraint is
denoted by the blue curve in Fig. \ref{plot1}.
In Fig. \ref{plot1} there are also two curves depicted with red
and green color. The green color curve corresponds to the
following choice of the free parameters $(\gamma,\beta)=(7.7\times
10^8,19312)$ and $\nu$ varies in the range $1\leq \nu \leq
23.0919$ with $\nu=1$ yielding a value $r\sim 0$ for $N=60$ and
$\nu=23.019$ yielding $r=0.07$. As it can be seen in Fig.
\ref{plot1}, the green curve is well fitted in the LIGO-Virgo
constraints depicted by the blue curve, for approximately $r\leq
5$. The red curve corresponds to the choices
$(\gamma,\beta)=(5,1.096)$ and $\nu$ again varies in the range
$(1,23.019)$. As it can be seen, only the values of $\nu$ in the
range $\nu$ $\epsilon$ $1\leq \nu \leq 21.8033$ yield a tensor
spectral index compatible with the LIGO-Virgo constraints. We need
to note that the overall behavior is qualitatively similar for
different choices of the values of the free parameters, and the
viability of the model is ensured for quite a large set of the
parameter values, but we omitted this analysis for brevity. The
behavior depicted in Fig. \ref{plot1} is characteristic for all
the cases omitted.

In conclusion, the power-law Einstein-Gauss-Bonnet model we
developed in this section yields a blue-tilted tensor spectral
index, which is compatible with the LIGO-Virgo data at frequencies
$20-84\,$Hz. It is interesting to discuss another interesting
possibility for successful phenomenological model building of
Einstein-Gauss-Bonnet model, this time in relation to the NANOGrav
astonishing result \cite{Arzoumanian:2020vkk,NANOGrav:2020spf}.
Recently the NANOGrav collaboration
\cite{Arzoumanian:2020vkk,NANOGrav:2020spf} reported a possible
detection of a stochastic gravitational wave background after
nearly thirteen years data set of the pulsar timing array.
Millisecond pulsars are among the most stable clocks existing in
nature. Therefore if several millisecond pulsars are observed
simultaneously, the slightly different timing residuals, or even
deformations on the arrival time if the signals emitted by them,
can originate by some intrinsic or even environmental noises, and
more interestingly by a stochastic gravitational waves background.
In the NANOGrav collaboration, the timing residuals originating
from an array of millisecond pulsars, has been analyzed coherently
to separate gravitational waves induced residuals from other
possible effects. Although a stochastic gravitational wave
background would surely be confirmed by NANOGrav if quadrupolar
spatial correlations are also confirmed, the possibility of having
a first stochastic gravitational background at frequencies $f\sim
10^{-8}$Hz is rather exciting. Now taking into account that Big
Bang Nucleosynthesis corresponds to $f\sim 10^{-11}$Hz
approximately, which furthermore corresponds to a Universe's
temperature $T\sim 0.1\,$MeV, or equivalently, redshift $z\sim 3
\times 10^8$, the NANOGrav frequencies correspond to an era during
the radiation domination era. These modes re-entered the horizon
during the radiation domination era and have a quite short
wavelength, although there exist quite shorter wavelengths probed
by the BBO or LISA collaboration. In any case, if the NANOGrav
data indeed correspond to a stochastic gravitational wave
background formed by inflationary gravitational waves, these modes
are inflationary tensor modes that re-entered the horizon during
the radiation domination era. The NANOGrav result indicated that
the strain amplitude of the corresponding signal is
$\mathcal{A}\sim 10^{-15}$ and the spectrum of the inflationary
gravitational waves would be $\Omega_{GW}\sim 10^{-8}$, which is a
quite large value when one considers single field inflation
models. In fact, due to the fact that single scalar field
inflation has a negative tensor spectral index, the NANOGrav
result cannot be explained by canonical single field inflation. An
exciting possibility was proposed in Refs.
\cite{Kuroyanagi:2020sfw,Vagnozzi:2020gtf}, and it was claimed
that an inflationary theoretical framework which predicts a
positive tensor spectral index with a relatively low reheating
temperature, can well describe the NANOGrav result. Thus in view
of this exciting possibility, the result we obtained in this work
can potentially describe the NANOGrav result, since our model
predicts a positive tensor spectral index. However a detailed
analysis of the gravitational wave power spectrum is required in
order to analyze all the different aspects and possibilities of
the phenomenological predictions of the model, and we shall
address these issues in a future work.

\section{Conclusions}

In this work we provided a self-consistent and simplified
Einstein-Gauss-Bonnet inflationary theoretical framework
compatible with the GW170817 event observational data.
Particularly, we surgically modified our previous work
\cite{Odintsov:2020sqy} in order to provide a simpler and more
functional theoretical inflationary framework. Our assumptions
were the slow-roll conditions, and in addition the conditions
(\ref{mainnewassumption}) and (\ref{mainconstraint2}), namely
$\kappa \frac{\xi '}{\xi''}\ll 1$ and
$\frac{4\kappa^4\xi'^2V}{3\xi''}\ll 1$. The first condition is
highly motivated by the slow-roll conditions, and specifically due
to the fact that the expression $\kappa \frac{\xi '}{\xi''}$
appears in the first slow-roll index. The second condition is
imposed solely in the cosmological system in order to extract
analytical results, thus it is the only condition imposed by hand.
As we demonstrated, the resulting theoretical framework is easy to
study analytically, and also provides us with relatively simpler
expressions for the slow-roll indices and the corresponding
observational indices in comparison to the ones of Ref.
\cite{Odintsov:2020sqy}. Accordingly, we used a simple power-law
model in order to investigate whether it is possible to obtain
viable inflationary models with this framework. As we
demonstrated, the power-law model is compatible with the latest
(2018) Planck data, and also the observational indices and the
slow-roll indices had particularly elegant and simple forms. A
notable feature of this power-law model is that for specific
values of the exponent in the scalar coupling function
$\xi(\phi)\sim \phi^{\nu}$, the tensor spectral index becomes
blue-tilted. We discussed this feature of the model in view of the
NANOGrav results, and as it was claimed in Refs.
\cite{Kuroyanagi:2020sfw,Vagnozzi:2020gtf} such blue-tilted
inflationary theory can explain the NANOGrav results, if the
signal indeed corresponds to a stochastic gravitational wave
background. However, this issue should be further discussed in
more detail, and this task stretches beyond the aims and scopes of
this paper. The detailed analysis which will be given elsewhere
should combine the generation of a successful inflationary era,
the generation of a viable dark energy era, and an appropriate
estimation of the primordial gravitational wave spectrum and
strain amplitude should be given. Also, with regard to the
predicted primordial gravitational wave spectrum, this should be
compared with the general relativistic one corresponding to a
standard scalar field inflationary model, with the consistency
relation $n_t=-r/8$ holding true. This detailed analysis is
deferred to a future work focused on this topic.

Let us note further that a significant improvement would be to
find a way to explicitly calculate $\ddot{\phi}$. This however is
not easy in the present context because the only way is by
differentiating the approximate relation (16), and if we do this,
and substitute in Eq. (5) one is lead to an inconsistency. So
truly it is not self consistent to use the approximate relation
and to our opinion this result should not be trusted, even if it
is validated. Unfortunately the lack of analyticity is an
obstacle. In addition, there are basically two different relations
that can define $\dot{\phi}$ by neglecting $\ddot{\phi}$, one is
Eq. (\ref{constraint1}) and the other one is (\ref{motion3}). Thus
if we explicitly differentiate the two different equations that
yield an expression for $\dot{\phi}$, things will be complicated
and there is doubt whether the results should be trustworthy. The
lack of analyticity and the presence of both the potential
$V(\phi)$ and of the scalar coupling function $\xi(\phi)$ makes
things significantly more complicated in comparison to the simple
scalar field case. The only way out might be in fact if one does
not neglect $\ddot{\phi}$ in Eq. (\ref{constraint1}), solve with
respect to $\dot{\phi}$ and substitute the results in Eq.
(\ref{motion3}) the scalar field equation of motion. In this case
basically the system of equations becomes much much more involved,
but we work toward this research line and the noteworthy results
will be reported in a future work. Another future perspective is
considering the possible trans-Planckian inconsistencies of the
models presented in this work. This might be useful due to the
large values of the scalar field which are of the order
$\mathcal{O}(20\,M_p)$, during inflation. It is expected that the
scalar field takes values of the order of Planck mass, but such
large values are notable to say the least, and should be
considered in view of the trans-Planckian aspects of the theory.
We hope to address these issues in the near future.

\end{document}